\documentclass[useAMS,usenatbib]{mn2e}

\usepackage{amsmath}
\usepackage{amssymb}
\usepackage{graphicx}
\usepackage{dcolumn}
\usepackage{bm}
\usepackage{ulem}

\title[On excess entropy and latent heat in white dwarfs]
{On excess entropy and latent heat in crystallizing white dwarfs}
\author[D. A. Baiko]{D. A. Baiko\thanks{E-mail:baiko@astro.ioffe.ru} \\
Ioffe Institute, Politekhnicheskaya 26, 194021 Saint Petersburg, Russia}

\begin{document}

\date{Accepted; Received ; in original form}

\pagerange{\pageref{firstpage}--\pageref{lastpage}} \pubyear{2014}

\maketitle

\label{firstpage}

\begin{abstract}
Based on the linear mixing approach, we calculate the latent heat 
for crystallizing fully-ionized $^{12}$C/$^{16}$O and 
$^{16}$O/$^{20}$Ne mixtures in white dwarf (WD) cores for two different 
parametrizations of the corrections to the linear-mixing energies and 
with account of ion quantum effects. We report noticeable 
composition-dependent deviations of the excess entropy in both 
directions from the standard value of 0.77 per ion. Within the same 
framework, we evaluate the excess entropy and released or absorbed heat 
accompanying the exsolution process in solidified WD layers. The 
inclusion of this effect is shown to be important for reliable 
interpretation of WD cooling data. We also analyze the latent heat of 
crystallizing eutectic $^{12}$C/$^{22}$Ne mixture, where we find a 
qualitative dependence of both the phase diagram and the latent heat 
behaviour on ion quantum effects. This may be important for the model 
with $^{22}$Ne distillation in cooling C/O/$^{22}$Ne WD proposed as a 
solution for the ultramassive WD multi-Gyr cooling anomaly. 
Astrophysical implications of our findings for crystallizing WD are 
discussed.
\end{abstract}




\section{Introduction}
Recent progress in white dwarf (WD) observations, made possible by the 
European Space Agency {\it Gaia} mission \citep[][]{G18}, have revealed 
certain deficienies in our understanding of evolution of these 
seemingly simple objects. In particular, \citet*{CCM19} have 
shown that a fraction of massive WD experiences a gigantic cooling delay 
of $\sim 8$ Gyr unaccounted for by standard models. 
Discoveries like this have tremendous impact on astrophysics, in which 
WD serve as cosmic chronometers, allowing one to gauge the age of 
various stellar populations \citep*[e.g.][]{FBB01,T+14}, and on 
strongly-coupled plasma physics. The latter is ultimately responsible 
for the multitude of processes in dense WD interiors, which manifest 
themselves in such spectacular ways, but cannot be emulated in 
terrestrial laboratories. 

The experimental advance has been accompanied by and has motivated
new theoretical work on various aspects of WD physics 
\citep*[see e.g.][for a recent review]{SBT22}. This includes, the 
ingenious distillation model proposed as a solution for the ultramassive
WD cooling anomaly \citep*[][]{BDS21}, recent work on diffusion in 
strongly-coupled plasma \citep*[e.g.][]{H+10,BY13,SM16,B+20_2,CBF22}, on 
classic binary and ternary phase diagrams 
\citep*[e.g.][]{B+20,CHC20,BD21}, on 
thermodynamics of fully ionized strongly-coupled {\it quantum} 
one-component 
plasma (OCP) of ions \citep[e.g.][]{BY19,BC22}. Based on the latter 
work, phase diagrams of quantum binary ionic mixtures have been 
analyzed using the linear mixing formalism by \citet{B22} (hereafter 
Paper I). 

In this letter, we aim to contribute to the WD cooling theory
by calculating the excess entropy and heat release or absorption 
accompanying crystallization and exsolution\footnote{Exsolution is a 
process, by which a single solid solution, stable at higher 
temperatures, at lower temperatures, unmixes into two separate solid 
phases. The process is well-known on Earth, where it can result in a 
formation of megascopic lamellar structures, e.g. 
\url{https://en.wikipedia.org/wiki/Perthite}} processes in 
$^{12}$C/$^{16}$O and $^{16}$O/$^{20}$Ne mixtures. In particular, we 
apply the same formalism as in Paper I 
to analyze the dependence of the excess entropy of crystallization and 
exsolution on energy parametrization and on mixture type and composition 
as well as to estimate the importance of ion quantum effects. We also 
study the phase diagram and the latent heat of crystallization for the 
eutectic $^{12}$C/$^{22}$Ne mixture at classic and quantum densities. 

The results are presented in section \ref{result}, including analytic
fits in subsection \ref{Fite}. A discussion of 
astrophysical implications is given in section \ref{discus}.

\section{Methods and results}
\label{result}
\begin{figure}
\begin{center}
\leavevmode
\includegraphics[bb=199 181 440 740, width=84mm]{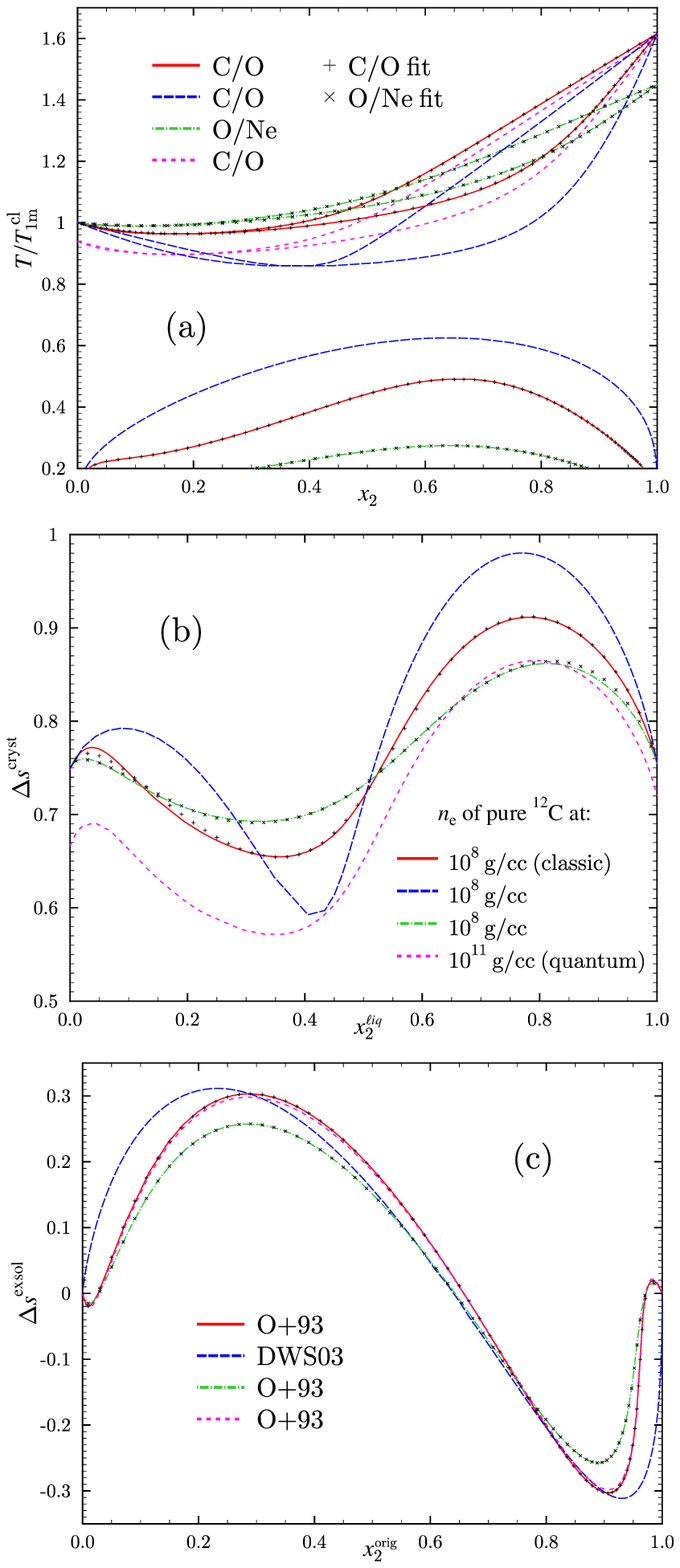}
\end{center}
\vspace{-0.4cm}
\caption[ ]{
(a) Phase diagrams of $^{12}$C/$^{16}$O (solid, dashed, short-dashed) 
and $^{16}$O/$^{20}$Ne (dot-dashed) mixtures at electron density
corresponding to pure carbon at $10^{8}$ g cm$^{-3}$ (solid, dashed, 
dot-dashed) and $10^{11}$ g cm$^{-3}$ (short-dashed) for the correction
to the linear-mixing energy given by the fit of \citet{O+93}
(solid, short-dashed, dot-dashed) and of \citet{DWS03} (dashed);
(b) excess entropy of crystallization as a function of the heavier
constituent number fraction in the liquid under the 
assumptions of panel (a) shown by the same curve types; 
(c) excess entropy of exsolution as a function of the heavier
constituent number fraction in the parent solid under the 
assumptions of panel (a) shown by the same curve types.
}
\label{d_e}
\end{figure}

In Fig.\ \ref{d_e}a, we remind a few results of Paper I. Shown are phase 
diagrams of $^{12}$C/$^{16}$O and $^{16}$O/$^{20}$Ne fully-ionized 
binary ionic mixtures under various assumptions. $Y$-axis is 
the temperature in units of classic melting temperature of the lighter 
component, $T_{\rm 1m}^{\rm cl}\equiv Z_1^{5/3} e^2/(a_{\rm e} 
\Gamma_{\rm m})$ [$k_{\rm B}\equiv 1$, 
$a_{\rm e} = (4 \pi n_{\rm e}/3)^{-1/3}$, $n_{\rm e}$ is the electron
density, and $\Gamma_{\rm m}=175.6$]. 
$X$-axis is the number fraction of 
the heavier component, $x_2=N_2/(N_1+N_2)$. Upper and lower portions of 
the plot display crystallization curves and miscibility gaps, 
respectively. Solid (red) lines correspond to $^{12}$C/$^{16}$O mixture 
at (essentially classic) electron density of pure carbon at 
$10^8$ g cm$^{-3}$, assuming correction to the linear-mixing 
energy given by the fit of \citet{O+93} and residual entropy of the 
solid equal to $S_{\rm mixZ}$ (see Paper I for details). Dashed (blue) 
curves use the fit of \citet{DWS03} 
and thus demonstrate sensitivity to a different energy correction 
parametrization. Short-dashed (magenta) curves assume a higher electron 
density corresponding to pure carbon at $10^{11}$ g cm$^{-3}$
and illustrate ion quantum effects\footnote{Mass densities 
$\sim 10^{11}$ g cm$^{-3}$ are typically not used in contemporary WD 
models but can be easily encountered in outer neutron star crust, where 
physics is very similar to that of WD interior. We use such an extreme 
density to illustrate maximum theoretically achievable strength of ion 
quantum effects in compact stars. Another possibility would be to 
consider helium at $\sim 10^7$ g cm$^{-3}$.} for $^{12}$C/$^{16}$O 
mixture. Finally, the dot-dashed (green) curves are the same as the 
solid ones but for $^{16}$O/$^{20}$Ne mixture.     

Our first goal is to calculate the excess entropy between phases shown
in Fig.\ \ref{d_e}a and any calorific effect associated with it. In 
order
to obtain the excess entropy, one needs to find the entropies of two
phases at their phase boundaries, i.e. at fraction values, for which
the phases are in equilibrium with each other at a given temperature.
Then take a difference between them. The entropies in question can be
written as $S=(E-F)/T$, where $E$ is the system energy and $F$ is its
Helmholtz free energy. Linear mixing expressions for $F$ in the liquid 
and solid phases are given e.g. in Paper I. Analagous formulae
for $E$ can be derived from them in a straightforward manner.  

In Fig.\ \ref{d_e}b, by curves of the same type, we show respective 
excess entropy per ion at crystallization 
$\Delta s^{\rm cryst} = [S/(N_1+N_2)]^{\ell{\sl iq}} 
- [S/(N_1+N_2)]^{{\sl so}\ell}$ as a function of $x_2$ in the liquid 
phase, $x_2^{\ell{\sl iq}}$. Upon multiplication by melting temperature, 
corresponding to this $x_2^{\ell{\sl iq}}$, it yields the respective 
latent heat of crystallization. For pure matter 
($x_2=0$ or 1), the excess entropy approaches its OCP value, which is 
a smooth function of density due to ion quantum effects 
\citep[cf.][]{BC22}. However, for intermediate fractions, 
$\Delta s^{\rm cryst}$ can noticeably deviate from this value. The 
amplitude of the deviation depends on mixture type, being larger 
for $^{12}$C/$^{16}$O mixture than for $^{16}$O/$^{20}$Ne. It is also 
sensitive to the assumed parametrization of the correction to the 
solid-state linear-mixing energy. For instance, in the classic 
situation, for the \citet{O+93} fit, the minimum excess entropy 
$\Delta s^{\rm cryst} \approx 0.65$ is at 
$x^{\ell{\sl iq}}_{\rm O} \approx 0.36$ and melting temperature 
$T_{\rm m} \approx 0.99 T_{\rm 1m}^{\rm cl}$. The maximum excess 
entropy $\Delta s^{\rm cryst} \approx 0.91$ is at 
$x^{\ell{\sl iq}}_{\rm O} \approx 0.78$ and 
$T_{\rm m} \approx 1.37 T_{\rm 1m}^{\rm cl}$. For the \citet{DWS03} fit,
the minimum excess entropy $\Delta s^{\rm cryst} \approx 0.59$ occurs at
$x^{\ell{\sl iq}}_{\rm O} \approx 0.41$ and 
$T_{\rm m} \approx 0.86 T_{\rm 1m}^{\rm cl}$, whereas the maximum 
excess entropy $\Delta s^{\rm cryst} \approx 0.98$ takes place at 
$x^{\ell{\sl iq}}_{\rm O} \approx 0.77$ and 
$T_{\rm m} \approx 1.29 T_{\rm 1m}^{\rm cl}$.

In Fig.\ \ref{d_e}c, we show the excess entropy accompanying the 
exsolution process. In this case, $\Delta s^{\rm exsol} = 
[S^{{\sl so}\ell}/(N_1+N_2)]^{\rm orig} 
- [S^{{\sl so}\ell}/(N_1+N_2)]^{\rm new}$, 
where indices `orig' and `new' refer to, respectively, the parent solid 
solution and the one nucleating from it upon cooling to the miscibility 
gap. The $x$-axis shows the heavier element fraction in the parent 
solution, $x^{\rm orig}_2$. The curve type has the same meaning as 
above. Note, that 
ion quantum effects cancel out when one determines the miscibility gap 
shape (see Paper I) but not the associated excess entropy.
Thus, solid and short-dashed curves coincide in the lower portion of 
Fig.\ \ref{d_e}a but not in Fig.\ \ref{d_e}c. All curves cross zero at 
$x^{\rm orig}_2 \sim 0.65$, which corresponds to maxima of their 
respective 
miscibility gaps. If $x^{\rm orig}_2$ in the original solution is to 
the left of 
the gap maximum, the exsolution results in a heat release of 
$T_{\rm mg} \Delta s^{\rm exsol}$ per each exsolved ion, where 
$T_{\rm mg}$ is the miscibility gap temperature at the parent 
$x^{\rm orig}_2$. In 
the opposite case, the exsolution will act as a heatsink absorbing 
$T_{\rm mg} |\Delta s^{\rm exsol}|$ as lighter solid nucleates. As the 
system cools down the solvus, the relative quantities of the original 
and new solids will be determined by the lever rule, and the heat 
release (or absorption) will continue with a decreasing $T_{\rm mg}$ 
and appropriately varying 
$|\Delta s^{\rm exsol}|$. In spite of the fact 
that $\Delta s^{\rm exsol}$ in Fig.\ \ref{d_e}c is similar for 
different models, the released or absorbed heat will be different 
due to dependence of $T_{\rm mg}$ at a given $x^{\rm orig}_2$ on 
ion sorts and on multicomponent plasma physics (i.e. on corrections to 
linear mixing).

Note, that in calculations of $\Delta s^{\rm cryst}$ and 
$\Delta s^{\rm exsol}$, the temperature-independent linear-mixing 
energy corrections in the solid state obviously cancel out. Yet, there 
is a difference between solid and dashed curves in Figs.\ \ref{d_e}b and 
\ref{d_e}c. This difference is due to different locations of the 
respective phase transitions and miscibility gaps in Fig.\ \ref{d_e}a.

\begin{figure}
\begin{center}
\leavevmode
\includegraphics[bb=199 367 440 741, width=84mm]{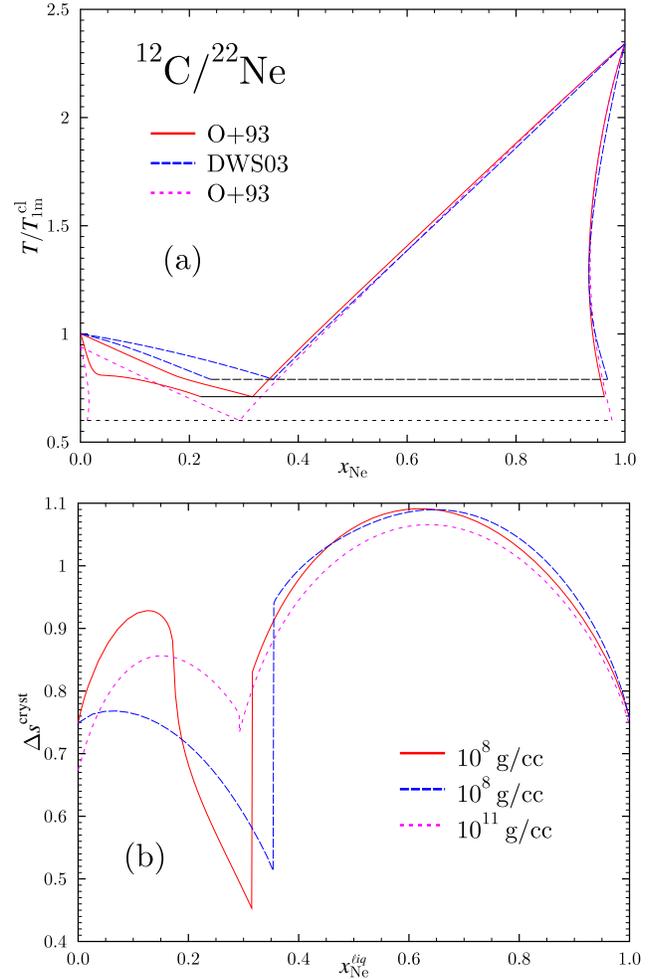}
\end{center}
\vspace{-0.4cm}
\caption[ ]{(a) Phase diagram of $^{12}$C/$^{22}$Ne mixture under 
the same assumptions as in Fig.\ \ref{d_e}. Thin horizontal 
lines show invariant region of the eutectic system; (b) excess entropy 
of crystallization for the system in panel (a).  
}
\label{eute}
\end{figure}

In Fig.\ \ref{eute}, we turn to $^{12}$C/$^{22}$Ne mixture, which is 
very similar to $^{12}$C/$^{20}$Ne mixture studied in Paper I. The 
liquid-solid coexistence curve under various assumptions is shown in 
Fig.\ \ref{eute}a. The low-$x_{\rm Ne}$ region of this diagram is 
somewhat representative of the low-$x_{\rm Ne}$ and low-$x_{\rm O}$ 
corner of the ternary C/O/$^{22}$Ne phase diagram, which is crucial 
for the distillation process proposed by \citet{BDS21}. We see a 
eutectic type phase diagram with solid (red) and short-dashed (magenta) 
curves corresponding to the correction fit of \citet{O+93} at the 
classic and quantum densities, respectively. Note, that, due to ion 
quantum effects, there is a qualitative change of the phase diagram 
structure at low $x_{\rm Ne}$ and, in particular, of neon depletion 
of the solid. The dashed (blue) curve is 
obtained using the correction fit of \citet{DWS03} at the same
classic density, and it varies only weakly under the action of quantum 
effects (cf.\ Paper I).       

In Fig.\ \ref{eute}b, by curves of the same type, we plot the excess 
entropy accompanying crystallization in this system. For pure systems, 
the excess
entropy approaches its OCP value (reduced to $\approx 0.67$ for quantum 
carbon). All curves have discontinuities at their eutectic points, which
are very strong for classic ions [e.g. from $\approx 0.45$ to 
$\approx 0.83$ for the \citet{O+93} fit]. Quantum effects reverse the 
jump sign and reduce its magnitude substantially for the \citet{O+93} 
fit but not for the \citet{DWS03} fit, for which quantum latent heat is
qualitatively similar to the classic one. At 
$x^{\ell{\sl iq}}_{\rm Ne} \sim$ 0.62--0.65, all curves reach 
maxima in the range of 1.07--1.09. Interestingly, the solid curve has 
another strong maximum of $\approx 0.93$ at a low  
$x^{\ell{\sl iq}}_{\rm Ne} \approx 0.13$.    

\subsection{Fit expressions}
\label{Fite}
To facilitate practical application of these results, we propose several 
analytic fits. For crystallizing classic $^{12}$C/$^{16}$O mixture, 
assuming the energy correction of \citet{O+93}, the phase diagram is 
described by functions 
$f_1\equiv T/T_{\rm 1m}^{\rm cl}(x^{\ell{\sl iq}}_{\rm O})$ and 
$f_2\equiv x^{{\sl so}\ell}_{\rm O}(x^{\ell{\sl iq}}_{\rm O})$
at the liquidus. Functions $f_3$ and $f_4$, respectively, represent 
the same quantities for crystallizing classic 
$^{16}$O/$^{20}$Ne mixture. For exsolving $^{12}$C/$^{16}$O mixture, 
assuming the energy correction of \citet{O+93}, function $f_5$ 
describes the rising segment of the solvus, i.e. 
$f_5\equiv T/T_{\rm 1m}^{\rm cl}(x^{\rm orig}_{\rm O})$ for 
$x^{\rm orig}_{\rm O} \lesssim 0.65$. Function $f_6$ yields the 
respective phase separation $x^{\rm new}_{\rm O}(x^{\rm orig}_{\rm O})$
with the range of validity of $f_5$ and $f_6$ limited by the condition 
$x^{\rm new}_{\rm O}\geq x^{\rm orig}_{\rm O}$. Using $f_5$ and $f_6$,
the whole solvus can be easily constructed. The respective functions 
for the $^{16}$O/$^{20}$Ne mixture are $f_7$ and $f_8$.

In the fitting formulae below, for brevity, the argument is denoted 
simply as $x$: 
\begin{eqnarray}
       f_1&=&\frac{T}{T_{\rm 1m}^{\rm cl}}(x^{\ell{\sl iq}}_{\rm O}) 
       = \frac{1+a_1 x+a_2 x^{2.1}+
       a_3 x^{3.3}+a_4 x^{4.5}}{1+a_5 x^{1.8}+a_6 x^{3.8}}~,
\nonumber \\
       f_2&=&x_{\rm O}^{{\sl so}\ell}(x^{\ell{\sl iq}}_{\rm O}) 
       = \frac{a_1 x+a_2 x^{1.5}+a_3 x^{1.8}+
       a_4 x^{7.5}}{1+(a_1+a_2+a_3+a_4-1) x^7}~,      
\nonumber \\
       f_3&=&\frac{T}{T_{\rm 1m}^{\rm cl}} (x^{\ell{\sl iq}}_{\rm Ne})
       = \frac{1+a_1 x^2+a_2 x^6}{1+
       a_3 x+a_4 x^2}~,
\nonumber \\
       f_4&=&x_{\rm Ne}^{{\sl so}\ell} (x^{\ell{\sl iq}}_{\rm Ne})
       = \frac{a_1 x+a_2 x^{1.6}+a_3 x^{4.5}+a_4 x^7}{1+
       (a_1+a_2+a_3+a_4-1) x^{6.6}}~,      
\nonumber \\
       f_{5,7}&=&\frac{T}{T_{\rm 1m}^{\rm cl}} (x^{\rm orig}_{\rm O, Ne})
       = \frac{a_1 x^{0.22}
       +a_2 x^{1.5}+a_3 x^{1.8}+a_4 x^{7.5}}{1+a_5 x^{1.1}+a_6 x^9}~,
\nonumber \\
       f_{6,8}&=&x_{\rm O, Ne}^{\rm new} (x^{\rm orig}_{\rm O, Ne})
       = 1-\frac{a_1 x+a_2 x^2
       +a_3 x^{3.5}}{1+a_4 x^{1.4}}~.      
\label{fiteq}
\end{eqnarray}
The coefficients $a_i$ are summarized in table \ref{fit}.

Once the phase boundaries are constructed, the excess entropies can be
found as $\Delta(E-F)/T$, where the energy and the free energy at the
boundaries are given
by the standard linear mixing formulae (e.g. Paper I). The accuracy of 
the fits (\ref{fiteq}) is sufficient to calculate relatively 
small entropies by subtracting $F$ from $E$, both dominated by 
relatively large electrostatic terms, and then, to 
subtract from each other entropies of the liquid and solid or of the 
parent and new solids. Data generated using the fits are shown by 
symbols in Figs.\ \ref{d_e}a--c.

\begin{table*}
\begin{center}
\begin{tabular}{ccccccc}
\hline
\hline
   & $a_1$ & $a_2$ & $a_3$ & $a_4$ & $a_5$ & $a_6$ \\
\hline
 $f_1$ &-0.45&0.557326&-0.287904&6.79381&-0.773978&4.49124  \\
 $f_2$ &0.55&1.41757&-0.640228&53.8103&&  \\
 $f_3$ &0.359282&-0.133598&0.164269&-0.318343&&  \\
 $f_4$ &0.8&0.747837&-4.98464&70.7369&&  \\
 $f_5$ &0.519262&-2.94837&8.01079&-7.18285&5.90273&4.02099  \\
 $f_6$ &1.80963&8.80669&107.8&155.816&&  \\
 $f_7$ &0.300237&-0.772378&3.05972&-7.99819&4.81564&-23.856 \\
 $f_8$ &1.52336&11.3999&84.2213&120.859&& \\
\hline
\end{tabular}
\end{center}
\caption{Fit coefficients}
\label{fit}
\end{table*}

\section{Discussion}
\label{discus}
Let us now discuss possible astrophysical implications of these 
findings. Since the work of \citet{S+00}, it has been typically assumed 
that the latent heat in crystallizing white dwarfs (WD) is equal to 
$0.77 T$ \citep[e.g.][]{C+19}. According to \citet{S+00}, earlier works 
used value of the latent heat $\sim 1 T$, and an introduction of the 
23\% smaller value resulted in a noticeable decrease of WD cooling age 
(by up to 0.4 Gyr). Depending on particular composition, a fraction of 
this effect may be reversed, if the larger values of the latent heat 
from Fig.\ \ref{d_e}b are used. 

In a recent work of \citet{J+21}, although specific value of the 
latent heat in an ionic mixture is not discussed explicitly (except 
a statement that it is similar to the standard OCP value), their 
Fig.\ 13 indicates that, 
depending on the numerical procedure, in the same object, the latent 
heat can be equal to $\sim 0.6 T$ or $\sim 0.95 T$. It is stated, that
the higher value is compensated, for cooling purposes, by a lower 
specific heat $c_p$. It follows, that in the approach of \citet{J+21}, 
a further cooling acceleration is predicted, compared with the standard 
latent heat case ($0.77 T$). Per their Fig.\ 15, 
$x^{\ell{\sl iq}}_{\rm O} \sim 0.7$, for which the solid (red) line in 
our Fig.\ \ref{d_e}b predicts the latent heat of $\sim 0.9 T$ and thus, 
by contrast, a cooling delay.

Conclusive study of the miscibility gap effects requires a separate 
work. Here, we shall limit ourselves to simplified estimates. Taking
as an example $x^{\ell{\sl iq}}_{\rm O} = 0.7$, we see (per solid 
lines in Fig.\ \ref{d_e}a) that the onset of crystallization corresponds 
to $T_{\rm m} \approx 1.27 T_{\rm 1m}^{\rm cl}$ and 
$x^{{\sl so}\ell}_{\rm O} \approx 0.84$. The miscibility gap
temperature at $x^{\rm orig}_{\rm O} = 0.84$ in the parent
solution is $T_{\rm mg} \approx 0.4 T_{\rm 1m}^{\rm cl}$, and the 
respective $\Delta s^{\rm exsol} \approx -0.25$ (Fig.\ \ref{d_e}c). 

From cooling models\footnote{
\url{https://www.astro.umontreal.ca/$\sim$bergeron/CoolingModels}} of 
\citet{B+20_3} (in what follows, we shall use thin hydrogen envelope 
models, thick hydrogen envelope models yield very similar results), one 
can estimate a change of the effective surface 
temperature, corresponding to a change of the internal temperature for 
WD of different masses. The exsolution onset occurs at the internal 
temperature $1.27/0.4 \approx 3.18$ times lower than the crystallization
onset temperature. Accordingly, one can deduce that the effective 
temperature at the exsolution onset is $\approx 1.3$, 1.5, and 1.75 
times lower than the effective temperature, at which $\approx 20 \%$ 
of WD mass is crystallized for 0.6 $M_\odot$, 0.8 $M_\odot$, and 
1 $M_\odot$ WD, respectively. 
This translates into a reduction of brightness by $\approx 1.15$, 1.8, 
and 2.45 stellar magnitudes, respectively. Comparing with Fig.\ 2 of 
\citet{T+19}, also cf. \citet{G18}, we see that the exsolution 
begins in a reasonably well populated area of the WD H-R diagram.
 
A useful measure of the process energetics is its effect on the cooling
age, which can be estimated as the ratio (denoted $\tau$ below) of 
the total process energy to the stellar luminosity
\citep[e.g.][]{I+97}. We shall compare 
these quantities for crystallization and exsolution under simplifying 
assumptions of constant temperature, constant fractions of 
constituents in the mixture, 
constant heat release/absorption per ion, 100\% of ions crystallizing,
and 100\% of carbon exsolving. Utilizing the same models as above, we 
deduce that the star at crystallization onset is 
$\eta \approx 3.65$, 7.75, and 15.3 times brighter than at the 
exsolution onset for 0.6 $M_\odot$, 0.8 $M_\odot$, and 1 $M_\odot$ WD, 
respectively. Then, 
$\tau^{\rm exsol}/\tau^{\rm cryst} \sim \eta \, x^{\rm orig}_{\rm C} 
T_{\rm mg} \Delta s^{\rm exsol}/(T_{\rm m} \Delta s^{\rm cryst}) 
\sim -0.05$, $-0.1$, and $-0.2$. Therefore, if crystallization (with 
the latent heat value of 
$0.9 T$) delays cooling by 1 Gyr, the exsolution will accelerate it (at 
a later stage) by about 50, 100, or 200 Myr depending on the stellar 
mass. 

Let us also consider an example of exsolution having the opposite 
effect. Suppose $x^{\ell{\sl iq}}_{\rm O} = 0.37$, so that 
$T_{\rm m} \approx T_{\rm 1m}^{\rm cl}$, parent 
$x^{{\sl so}\ell}_{\rm O} \approx 0.43$, 
$\Delta s^{\rm cryst} \approx 0.66$,
$T_{\rm mg} \approx 0.4 T_{\rm 1m}^{\rm cl}$, and 
$\Delta s^{\rm exsol} \approx +0.25$. The star at exsolution is 
$\eta \approx 2.7$, 6.2, and 8.9 times less luminous than at 
crystallization (for the same masses as above), and
$\tau^{\rm exsol}/\tau^{\rm cryst} \sim 0.2$, 0.5, and 0.8 
(assuming all oxygen 
exsolving). Thus, if due to the latent heat, crystallization delays 
cooling by 1 Gyr, the exsolution will add another 200, 500, or 800 Myr    
for 0.6 $M_\odot$, 0.8 $M_\odot$, or 1 $M_\odot$ WD, 
respectively.

In view of these estimates, we tend to conclude that the exsolution 
process should be taken into account quantitatively to improve 
interpretation of WD cooling data. Note also, that these thermal 
effects may be augmented by gravitational energy release associated 
with separation of lighter and heavier solids and lamellar structure 
formation in a stellar gravitational field (Paper I).

The phase diagram in Fig.\ \ref{eute}a predicts that if 
$x^{\ell{\sl iq}}_{\rm Ne}$ is to the left of the eutectic point (the 
V-type feature at $x_{\rm Ne} \sim$ 0.3--0.35), the crystal that 
forms will be depleted of $^{22}$Ne and therefore will be buoyant in the
surrounding neon-enriched fluid. This is the basis of the distillation 
model proposed by \citet{BDS21}. Clearly, the depletion becomes 
stronger and buoyancy is enhanced, as the solid (red) curve transitions 
into the short-dashed (magenta) one\footnote{Since $10^{11}$ g cm$^{-3}$ 
is too high a density for WD, a complete transition to the 
magenta curve does not happen.} with increase of the mass density 
(and WD mass), i.e. with amplification of ion quantum effect. A possible 
caveat may be related to the latent heat release, which warms up the 
surrounding fluid and reduces crystal buoyancy in it. If real, it is 
expected to be more pronounced for classic ions, for which there is a 
strong maximum of the latent heat at low $x^{\ell{\sl iq}}_{\rm Ne}$ 
(solid line in Fig.\ \ref{eute}b).

\section*{Acknowledgments}
The author is grateful to the anonymous referee for useful suggestions.
This work was supported by RSF grant 19-12-00133-P.

\section*{Data Availability}
The data underlying this article will be shared on reasonable 
request to the author.

\end{document}